\documentclass[aps,floats,twocolumn,superscriptaddress,prl,showpacs]{revtex4}
\usepackage{amsmath}

\usepackage{epsfig}
\usepackage{graphicx}
\usepackage{bm}

\newcommand{\beq}{\begin{equation}}
\newcommand{\eeq}{\end{equation}}
\newcommand{\beqa}{\begin{eqnarray}}
\newcommand{\eeqa}{\end{eqnarray}}
\def\gapp{\lower.35em\hbox{$\stackrel{\textstyle>}{\sim}$}}
\def\lapp{\lower.35em\hbox{$\stackrel{\textstyle<}{\sim}$}}

\begin{document}
\bibliographystyle{naturemag}
%


\title{Elastic Gauge Fields in Weyl Semimetals}

\author{Alberto Cortijo}
\affiliation{Instituto de Ciencia de Materiales de Madrid,\\
CSIC, Cantoblanco; 28049 Madrid, Spain.}

\author{ Yago Ferreir{\'o}s}
\affiliation{Instituto de Ciencia de Materiales de Madrid,\\
CSIC, Cantoblanco; 28049 Madrid, Spain.}
\author{ Karl Landsteiner}
\affiliation{Instituto de F\'isica Te\'orica UAM/CSIC, \\
Nicol\'as Cabrera 13-15, Cantoblanco, 28049 Madrid, Spain }

\author{Mar\'{\i}a A. H. Vozmediano}
\affiliation{Instituto de Ciencia de Materiales de Madrid,\\
CSIC, Cantoblanco; 28049 Madrid, Spain.}

\date{\today}
\begin{abstract}
We show that, as it happens in graphene,  elastic deformations couple to the electronic degrees of freedom as pseudo gauge fields in Weyl semimetals. We derive the form of the elastic gauge fields in  a tight-binding model hosting Weyl nodes and see that this vector electron-phonon coupling is chiral, providing an example of axial gauge fields in three dimensions.  As an example of the new response functions that arise associated to these elastic gauge fields,  we derive a non-zero phonon  Hall viscosity for the neutral system at zero temperature. The axial nature of the fields provides a test of the chiral anomaly in high energy with three axial vector couplings.
\end{abstract}
%
%
%
%
 \maketitle

\emph{Introduction.--} 
The occurrence of Weyl fermions (massless Dirac fermions of definite chirality) in condensed matter has come always with unexpected phenomena and new physics.  
Although having a long tradition \cite{Vo03}, the best examples so far arose in one spacial dimension (Luttinger liquids) \cite{G04} or in two (Graphene  \cite{CGP09} and the surface of three dimensional topological insulators \cite{QZ11}). Charged massless fermions are
particularly interesting in three dimensions:  They do not have counterparts in particle physics and they experience the chiral anomaly \cite{Adler69,BJ69,NN81,Grushin12,GT13} and its related physical responses. 

The Dirac equation comes from the existence of band crossings, ``Fermi points" in the dispersion relation and subsequent low energy expansion around them. Coming from a lattice, these Weyl fermions must always arise in pairs of oposite chirality -- or helicity -- according to the Nielsen-Ninomiya theorem \cite{NN81}. Dirac semimetals \cite{LJetal15} have the crossing points at the gamma point   of the Brillouin zone and the contribution from the two oposite chiralities cancel the anomaly related responses. The importance of the so--called Weyl semimetals is that, as happens in graphene, the two chiral partners sit at non equivalent points in momentum space and the physics of anomalies is present in full glory. This is why the recent experimental discovery of Weyl semimetals 
(WSM) \cite{HXetalNat15,XBetalSc15,XAetal15,Huetal15,BEetal15} is attracting so much excitement \cite{Joh15}. 
 
%
WSM have been named ``the 3D graphene". One of the most exotic and fruitful aspects of graphene has arisen from the demonstration that elastic lattice deformations couple to its electronic excitations in the form of fictitious gauge fields \cite{VKG10}. This fact, first deduced in a tight--binding model \cite{SA02b} was soon recognized to arise from very general symmetry considerations \cite{Ma07,MJSV13}. The experimental observation of the predicted Landau levels associated to the elastic magnetic fields \cite{GKG10,LBetal10}, have given rise to a whole new field of research called ``straintronics".

%
\begin{figure*}[!ht]
\includegraphics[width=\textwidth]{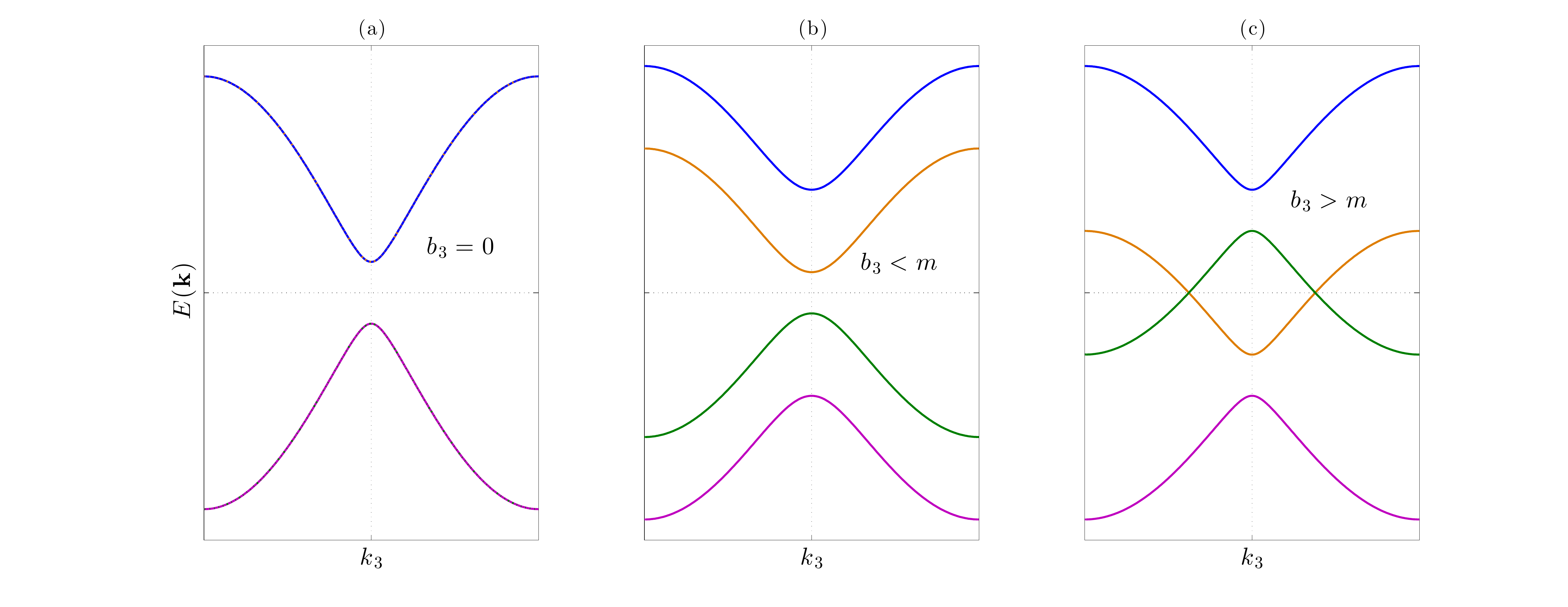}
\caption{(Color online.) Evolution of the band structure of the model given in eq. \eqref{Hzerob}as a function of the parameter $b$. For $b_{3}=0$ the spectrum consists in two pairs of degenerate bands due to time reversal symmetry, (a). When $0<b_{3}<m$ the band degeneracy breaks down and a high energy sector differentiates from a low energy sector, but the system is still gapfull, (b). When $b_{3}>m$, the low energy bands cross each other at two definite points in the Brillouin zone. At sufficiently low energies, the system consists in two pairs of Weyl fermions with opposite chirality, (c).}
\label{Hbbands}
\end{figure*}
In what follows we will show the presence of elastic gauge fields in WSM.
We first make a microscopic derivation  starting from a tight-binding description \cite{SA02b} and taking  the continuum limit around the Weyl points. As a physical consequence, 
we will show the presence of an anomalous phonon Hall viscosity in the WSM with time reversal symmetry ($\mathcal{T}$) broken. Similar to the Hall conductivity, the Hall viscosity can be used to classify topologically non-trivial states of matter  \cite{ASZ95,TV07,Read09,HLF11,BGR12,HS12}. We will see that the elastic gauge fields provide a new mechanism for generating Hall viscosity not previously studied in the literature. Due the chiral nature of the coupling between Weyl fermions and elastic degrees of freedom, this new  coupling provides an example of axial vector-fermion interaction with no analogue in high energy physics, and paves the way for studying the consequences of such couplings in a more general context. 
\\
\emph{Elastic gauge fields in a model for WSM.--} To illustrate how emergent vector fields associated to elasticity appear in a WSM phase we can consider the following simple model of s-, and p-like electrons hopping in a cubic lattice and chirally coupled to an on-site constant vector field $\bm{b}$ \cite{VF13,SHR15}
\begin{eqnarray}
H_{0}&=&\sum_{i,j}c^{+}_{i}\left(i t\alpha_{j}-r\hat\beta\right)c_{i+ j}+(m+3r)\sum_{i}c^{+}_{i}\hat\beta c_{i}+\nonumber\\
&+&\sum_{i,l} b_{l} c^{+}_{i}\alpha_{l}\gamma_{5}c_{i}+ h. c.,\label{Hzerob}
\end{eqnarray}
where $i$ labels the position $\mathbf{R}_{i}$ and $j$ labels the six next nearest neighbors $\mathbf{a}_{j}$ of length $a$ in the cubic lattice. The matrices $\alpha_{i}$ and $\hat\beta$ are the standard Dirac matrices. In the unstrained situation we will set all the hopping terms $t$ equal for simplicity. The parameters $t$, $r$, and $m$, represent, in a tight-binding description, the hopping matrix elements between $s$ and $p$ states, hopping between the same kind of states, and the difference of on-site energies between $s$ and $p$ states, respectively. The vector field $\bm{b}$ breaks $\mathcal{T}$ as well as the cubic lattice symmetry and thus the SO$(3)$ rotational symmetry in the continuum description. Without loss of generality, we will choose the vector field $\bm{b}$ to point along the OZ direction.
The model of eq. \eqref{Hzerob} with $b_{3}=0$ is the standard model to exemplify the transition from a trivial to a topological insulating phase as a function of the parameters $m$ and $r$. The region $0>m>-2r$ corresponds to a topological insulating phase and the long wavelength limit  around the $\Gamma$ point ($\bm{k}=0$) corresponds to an isotropic massive Dirac system.
As it can be seen in Fig.(\ref{Hbbands}) the WSM phase appears when $b_{3}>m$. The spectrum of the Hamiltonian in eq.(\ref{Hzerob}) consists of two bands crossing at two Fermi points in the BZ  and two bands at higher and lower energies (Fig.(\ref{Hbbands},c)). Let us choose the following representation for the Dirac matrices: $\alpha_{1}=\tau_{0}\sigma_{1}$, $\alpha_{2}=\tau_{0}\sigma_{2}$, $\alpha_{3}=\tau_{1}\sigma_{3}$, $\hat{\beta}=\tau_{3}\sigma_{3}$, $\gamma_{5}=\tau_{1}\sigma_{0}$, so $\alpha_{3}\gamma_{5}=\tau_{0}\sigma_{3}$. Fourier transforming (\ref{Hzerob}) we obtain:
\begin{widetext}
\begin{eqnarray}
\mathcal{H}_{0}(\bm{k})=\left(\begin{array}{cc}
t\sum_{s}\sigma_{s}\sin(k_{s}a)+(b_{3}+m(\bm{k}))\sigma_{3} & t\sin(k_{3}a)\sigma_{3}\\
t\sin(k_{3}a)\sigma_{3} & t\sum_{s}\sigma_{s}\sin(k_{s}a)+(b_{3}-m(\bm{k}))\sigma_{3}\end{array}\right),
\end{eqnarray} 
with $s=(1,2)$, and $m(\bm{k})=m+3r-r\sum_{j}\cos(k_{j}a)$. The four-component wavefunction can be written in two-component blocks $\left(\phi_{\bm{k}},\psi_{\bm{k}}\right)$. For energies $E\ll m+b_{3}$  we can write $\phi_{\bm{k}}\simeq-\frac{v k_{3}}{m+b_{3}}\psi_{\bm{k}}$. Projecting out the high energy sector represented by $\phi_{\bm{k}}$ and expanding around the $\bm{k}=0$ point, we find the following effective two-band model in the continuum ($v=ta$):
\begin{equation}
H_{eff}=\sum_{\bm{k}}\psi^{+}_{\bm{k}}\left(v\bm{\sigma}_{\perp}\cdot\bm{k}_{\perp}+\frac{1}{m+b_{3}}(b^{2}_{3}-m^{2}-v^{2}k_{3}^2)\sigma_{3}\right)\psi_{\bm{k}}.\label{Heff}
\end{equation} 
\end{widetext}
The existence of Weyl points $\bm{\lambda}$ is met when for $k_{1}=k_{2}=0$,  the equation $b^{2}_{3}-m^{2}-v^2k_{3}^2=0$ has real solutions. In this case, the two Weyl points are located at $\bm{\lambda}_{\pm}=(0,0,\pm\sqrt{\frac{b^2_{3}-m^2}{v^2}})$. Expanding now around these two points $\bm{k}\simeq \bm{\lambda}_{\pm}+\delta\bm{k}$, the low energy effective  Hamiltonian takes the form of two massless three dimensional Dirac fermions $\psi_{\pm}$ separated by the vector $\bm{\lambda}_+-\bm{\lambda}_-$ in momentum space:
\begin{equation}
H_{W}=\sum_{\delta\bm{k}}\psi^{+}_{\pm,\delta\bm{k}}\left(v\bm{\sigma}\cdot\delta\bm{k}_{\perp}\mp v_{3}\delta k_{3}\sigma_{3}\right)\psi_{\pm,\delta\bm{k}},\label{Weyleff}
\end{equation}
with $v_{3}=2v\sqrt{\frac{b_3-m}{b_3+m}}$.

Now we will apply strain to the original tight binding Hamiltonian and find the modifications it induces in the low energy Hamiltonian \eqref{Weyleff}. The strain tensor $u_{ij}$ enters in the tight binding approach through the change of the hopping parameters $t$ when the lattice is distorted. In the model of eq. \eqref{Hzerob}, there are two types of corrections to $t$.  One, similar to that arising in graphene \cite{VKG10}, is due to the change in the bond length. It is isotropic and exists for all orbitals:
\begin{equation}
r\rightarrow r_{j}\simeq r(1-\beta u_{jj}),
\end{equation}
where $\beta$ is the Gr\"{u}neisen parameter of the model. 
The second  contribution affects the hopping between different orbitals and is associated to a rotation with respect to the reference frames of neighbouring atoms as described in \cite{SHR15}. Following this reference, the changes for $t_{j}$ are: 
\beq
t\alpha_{j}\rightarrow t(1-\beta u_{jj})\alpha_{j}+t\beta\sum_{j'\neq j}u_{jj'}\alpha_{j'}.
\eeq
Inserting these modifications in the original Hamiltonian (\ref{Hzerob}), we can define the strained Hamiltonian as the sum of the original Hamiltonian $H_{0}$ and the strain dependent part $H[u_{ij}]$. Projecting out the high energy sector and expanding around the two nodal points $\bm{\lambda}_{\pm}$, the strain dependent Hamiltonian part takes the form
\begin{eqnarray}
H^{W}_{eff}[u]_{\pm}=\pm \beta\sqrt{b_3^2-m^2}\sum_{\bm{k},s=1,2} u_{3s}\psi^{+}_{\pm,\bm{k}}\sigma_{s}\psi_{\pm,\bm{k}}+\nonumber\\
+\beta\sum_{\bm{k}}\left(2(b_3-m)u_{33}-r \sum_ju_{jj}\right)\psi^{+}_{\pm,\bm{k}}\sigma_{3}\psi_{\pm,\bm{k}}.
\end{eqnarray}
We have found that, around the two nodal points, strain couples to the low energy electronic sector as a vector field:
\begin{eqnarray}
A^{el}_{1}=\beta \sqrt{b^{2}_{3}-m^{2}} u_{31},\nonumber
\\
A^{el}_{2}=\beta\sqrt{b^{2}_{3}-m^{2}} u_{32},\nonumber
\\
A^{el}_{3}=2\beta(b_3-m)u_{33}-\beta r\sum_ju_{jj}.
\label{3Delasticfield}
\end{eqnarray}
This is the first main result of this Letter. 

The low energy effective action in the continuum limit around the Weyl nodes ($\bm{\lambda}_{\pm}$) is thus given by 
\beq
H_{W}=\sum_{\delta\bm{k}}\psi^{+}_{\pm,\delta\bm{k}}\left(\bm{\sigma}(v\delta\bm{k}_{\perp}\pm\bm{A}^{el}_\perp)\mp 
(v_{3}\delta k_{3}\pm A_3^{el})\sigma_{3}\right)\psi_{\pm,\delta\bm{k}}.\label{FinalHam}
\eeq
Similarly to what happens in graphene or other two dimensional systems, strain couples to electrons as a chiral vector field i. e. it couples with opposite signs to the electronic excitations around the two Weyl nodes.
The specific form  of \eqref{3Delasticfield} is due to the choice of the vector ${\bf b}$ pointing along the $OZ$ axis. 
\\
\emph{Hall viscosity.--} As a physical consequence of the presence of the elastic gauge fields, we will next show that WSM have an intrinsic Hall viscosity. In visco-elastic systems the viscosity tensor is defined as the transport coefficient relating the stress tensor $\tau_{ij}$ and the \emph{time derivative} of the strain tensor $u_{rs}$, $\tau_{ij}=\eta_{ijrs}\dot{u}_{rs}$. The antisymmetric part of $\eta_{ijrs}$ is a dissipationless coefficient allowed only when $\mathcal{T}$ is broken. In three  dimensions rotational symmetry must also be broken to get a nonvanishing Hall viscosity. For axially symmetric systems with broken  $\mathcal{T}$ there are two independent components of the Hall viscosity tensor that can be chosen $\eta_{3231}$ and $\eta_{1112}$ \cite{ASZ95}. The Hall viscosity was first defined as an intrinsic property of the quantum fluid. When a topologically non-trivial electronic fluid is coupled to the crystal environment it will induce a Hall viscosity term in the elastic free energy of phonons \cite{BCQ12}. The electronic contribution to the field theory for the elastic degrees of freedom can be obtained by integrating out the electronic fields in (\ref{FinalHam}). The effective action will contain the following term (in units $\hbar=1$):
\begin{eqnarray}
&&\Gamma_{H}[u]=\frac{1}{48\pi^{2}}\int d^4 x \epsilon^{\mu\nu\rho\sigma}\lambda_{\mu}A^{el}_{\nu}\partial_{\rho}A^{el}_{\sigma}=\nonumber\\
&=&\frac{\beta^{2}}{48\pi^{2}a^{3}}\left(\frac{b^{2}_{3}-m^{2}}{t^{2}}\right)^{\frac{3}{2}}\int d^4 x
\left(u_{31}\dot{u}_{32}-u_{32}\dot{u}_{31}\right).
\label{CSWSM}
\end{eqnarray}
From this expression, we can easily read the coefficient $\eta_H=\eta_{3231}$ of Hall viscosity coming from the presence of the elastic gauge fields:
\begin{equation}
\eta_{H}=\frac{\beta^{2}}{24\pi^{2}}\frac{1}{a^{3}}\Big(\frac{b^{2}_{3}-m^{2}}{t^{2}}\Big)^{\frac{3}{2}}.
\label{HVWSM}
\end{equation}
This is the second main result of this letter: In the simple model considered, both $\mathcal{T}$ and rotational invariance are broken by the presence of the constant vector $\bm{b}$ giving rise to a Hall viscosity through the elastic vector fields. This response is rooted on the topological nature of the material and is universal in the sense that it is directly related to the Hall conductivity. For a general 3D Weyl semimetal  breaking time reversal symmetry, the anomalous hall effect is characterized by a momentum space vector  called the Chern vector. In our model, this is the vector $\lambda_\mu$ separating the two Weyl nodes in momentum space. The anomalous hall conductivity is given by the expression:
$
\sigma_{ij} = \frac{e^2}{ 2\pi c} \epsilon_{ijk}\lambda_k,
\label{3DHE}
$
coming from a 3D Chern Simons term of the form
\beq
S_{CS}\sim \nu_H\int d^4 x \lambda_i\epsilon^{ijkl}A_j\partial_k A_l,
\label{3DCS}
\eeq
where $\nu_H$ is the 3D Hall conductivity. It is easy to recognize the first term of eq. \eqref{CSWSM} as the  Chern Simons term associated to the elastic gauge fields. As a rule, what we have shown is that, any Hall system supporting elastic gauge fields will automatically present a Hall viscosity response.
\\
\emph{Discussion.--} As a proof of concept, we have shown that WSMs couple to elasticity through chiral vector fields by using a minimal tight-binding lattice model. This fact is not tied to the breakdown of time reversal symmetry, although we have used a model where the Weyl points appear by breaking $\mathcal{T}$ (a system with broken time reversal symmetry has  been reported in\cite{BEetal15}). A necessary condition (albeit it might not be sufficient)  for having such elastic gauge fields (technically, to have a vector representation of the elastic degrees of freedom at the Weyl point) is to have the Weyl points sitting at non-equivalent points of the Brillouin zone (what excludes the $\Gamma$ point) \cite{Ma07}. The presence of Weyl points is compatible with $\mathcal{T}$ if the pair of Weyl nodes are related  by inversion symmetry $\mathcal{I}$ \cite{Manes12}. This implies that these elastic gauge fields will appear in most of the $\mathcal{T}$-invariant systems displaying Weyl nodes, implying the generality of this phenomenon. 

As a direct consequence of this chiral vector coupling between elasticity an the electronic degrees of freedom,  a new type of Hall viscosity arises in WSMs. In three dimensions Hall viscosity have been discussed in the literature associated to two instances only:  The topological insulator phase with $\mathcal{T}$ broken and a WSM system in the presence of torsion \cite{SHR15,SW14}. The elastic gauge fields coupled to a topologically non--trivial system defines a third mechanism that will act on topologically non trivial crystals supporting elastic vector fields.

Several aspects of the viscoelastic response of lattice topological crystals have been recently analyzed in  \cite{SHR15}. Our new contribution to the Hall viscosity, although  not explicitly discussed, could certainly have been worked out  as a part of their general analysis. The examples chosen there having the Weyl nodes at the gamma point prevented them from finding the elastic gauge fields.  The coupling giving rise to the  Hall viscosity in
that reference is linear in momentum and corresponds to the standard phonon 
viscosity found in the hydrodynamic approach \cite{TV07}. In contrast, the elastic gauge field term described in our work couples directly to the fermionic current and is of lowest order in a derivative expansion. 

Another important aspect of the present analysis arises in the connection of the new term with the chiral anomaly. As we discussed, the elastic vector fields are chiral in the sense that they couple with oposite signs to the two chiralities. Moreover, the field $\lambda_\mu$ is also an axial field what implies that the coefficient in eq. \eqref{CSWSM} is associated to the triangular graph with three axial vertices (AAA). This triangular graph  has an additional symmetry factor of 
1/3 compared to the usual one (one axial and two vector vertices AVV). We would like to emphasize that this gives rise to the exciting possibility to test the AAA anomaly in a condensed matter context. In contrast it is generally believed that this type of anomaly does not lead to physical 
consequences in high energy physics \cite{Sch13}.


The existence of elastic gauge fields in WSMs extends the field of ''straintronics" to three-dimensional materials and paves the path for the study of a plethora of emergent phenomena. Notice that Weyl points are not an exclusive property of the dispersion relation of  electronic systems. They have also been described in three dimensional photonic systems \cite{LFJ13,LWDetal15} what allows to envisage the extension of "straintronics" to photonic media by controlling the Weyl nodes with deformations through these elastic gauge fields. 
\\
\emph{Acknowledgments.--} We thank Carlos Hoyos for enlightening discussions on the Hall viscosity and Juan Ma\~nes for comments on the elastic gauge fields.  Special thanks go also to Jos\'e Silva-Guill\'en for help with the figures. This research was supported in part by the Spanish MECD grants FIS2011-23713, PIB2010BZ-00512, the  European Union structural funds and the Comunidad de Madrid MAD2D-CM Program (S2013/MIT-3007), by the National Science Foundation under Grant No. NSF PHY11-25915, and by the European Union Seventh Framework Programme under grant agreement no. 604391 Graphene Flagship, FPA2012-32828 and by the
Centro de Excelencia Severo Ochoa Programme under grant SEV-2012-0249.


\bibliography{HV2}

\end{document}